\documentclass[10pt,journal,compsoc]{./sty/IEEEtran}
\usepackage[utf8]{inputenc}
\usepackage{multirow}
\usepackage{multicol}
\usepackage[pdftex]{graphicx}
\usepackage{amsmath}
\usepackage{textcomp}
\usepackage{xcolor}
\usepackage{float}
\usepackage{hyperref}
\hypersetup{
   colorlinks=true,
   linkcolor=black,
   anchorcolor=black,
   citecolor=black,
   filecolor=black,
   urlcolor=black
}

\newcommand*{\RR}[1]{\textcolor{black}{#1}}
\newcommand*{\TR}[1]{\textcolor{black}{#1}}
\newcommand*{\HL}[1]{\textcolor{black}{#1}}
\newcommand*{\SK}[1]{\textcolor{black}{#1}}

\title{A Novel Policy for Pre-trained \\Deep Reinforcement Learning for \\Speech Emotion Recognition }

\author{%
 	\IEEEauthorblockN{Thejan Rajapakshe\IEEEauthorrefmark{1}\IEEEauthorrefmark{2},
 		Rajib Rana\IEEEauthorrefmark{2}, 
 		Sara Khalifa\IEEEauthorrefmark{3} ,
 		Bj\"{o}rn W.\ Schuller\IEEEauthorrefmark{4}\IEEEauthorrefmark{6},
 		Jiajun Liu\IEEEauthorrefmark{3} 
 	}
 	\IEEEauthorblockA{ \\
 		\IEEEauthorrefmark{2}University of Southern Queensland, Australia \\
 		\IEEEauthorrefmark{3}Distributed Sensing Systems Group, Data61, CSIRO Australia \\
 		\IEEEauthorrefmark{4} GLAM -- Group on Language, Audio \& Music, Imperial College London, UK \\
 		\IEEEauthorrefmark{6}ZD.B Chair of Embedded Intelligence for Health Care \& Wellbeing, University of Augsburg, Germany \\
 		\IEEEauthorrefmark{1} Thejan.Rajapakshe@usq.edu.au
 	}
}%

\date{}

\begin{document}

\maketitle

\begin{abstract}
	Reinforcement Learning (RL) is a semi-supervised learning paradigm where an agent learns by interacting with an environment. Deep learning in combination with RL provides an efficient method to learn how to interact with the environment called Deep Reinforcement Learning (deep RL). Deep RL has gained tremendous success in gaming -- such as AlphaGo, but its potential has rarely being explored for challenging tasks like Speech Emotion Recognition (SER). Deep RL being used for SER can potentially improve the performance of an automated call centre agent by dynamically learning emotion-aware responses to customer queries. While the policy employed by the RL agent plays a major role in action selection, there is no current RL policy tailored for SER. In addition, an extended learning period is a general challenge for deep RL,  which can impact the speed of learning for SER. Therefore, in this paper, we introduce a novel policy -- the ``Zeta policy'' which is tailored for SER and apply pre-training in deep RL to achieve a faster learning rate. Pre-training with a cross dataset was also studied to discover the feasibility of pre-training the RL agent with a similar dataset in a scenario where real environmental data is not available. The IEMOCAP and SAVEE datasets were used for the evaluation with the problem being to recognise the four emotions happy, sad, angry, and neutral in the utterances provided. The experimental results show that the proposed ``Zeta policy'' performs better than existing policies. They also support that pre-training can reduce the training time and is robust to a cross-corpus scenario.    

\end{abstract}


\section{Introduction}

Reinforcement Learning (RL), a semi-supervised machine learning technique, allows an agent to take actions and interact with an environment to maximise the total rewards. RL\textquotesingle s main popularity comes from its super-human performance in solving some games like AlphaGo \cite{Silver2016MasteringSearch} and AlphaStar \cite{Vinyals2019GrandmasterLearning}.

\SK{RL has been also employed for audio-based applications showing its potential for audio enhancement \cite{Shen2019ReinforcementRecognition, Fakoor2018ReinforcementQuality}, automatic speech recognition \cite{Chung2020Semi-supervisedLearning}, and spoken dialogue systems \cite{Singh1999ReinforcementSystems., Paek2006ReinforcementDeployment}. The potential of using RL for speech emotion recognition has been recently demonstrated for robot applications where the robot can detect an unsafe situation earlier given a human utterance \cite{Lakomkin2018EmoRL:Learning}. An emotion detection agent was trained to achieve an accuracy-latency trade-off by punishing wrong classifications as well as too late predictions through the reward function. Motivated by this recent study,} we employ RL for speech emotion recognition where a potential application of the proposed system could be an intelligent call centre agent, learning over time how to communicate with human customers in an emotionally intelligent way.
We consider the speech utterance as ``state'' and the classified emotion as ``action''. We consider a correct classification as a positive reward and negative reward, otherwise.

RL widely uses Q-learning, a simple, yet quite powerful algorithm to create a state-action mapping, namely Q-table, for the agent. This, however, is intractable for a large or continuous state and/or action space. First, the amount of memory required to save and update that table would increase as the number of states increases.
Second, the amount of time required to explore each state to create the required Q-table would be unrealistic.
To address these issues, the deep Q-Learning algorithm uses a neural network to approximate a Q-value function. Deep Q-Learning is an important algorithm enabling deep Reinforcement Learning (deep RL)~\cite{Fan2019AQ-Learning}. 

The standard deep Q-learning algorithm employs a stochastic exploration strategy called $\epsilon$-greedy, which follows a greedy policy according to the current Q-value estimate and chooses a random action with probability $\epsilon$. Since the application of RL in speech emotion recognition (SER) is mostly unexplored, there is not enough evidence if the $\epsilon$-greedy policy is best suited for SER. In this article, we investigate the feasibility of a tailored policy for SER. 
\TR{We propose a  Zeta-policy}
\footnote{Code: \url{https://github.com/jayaneetha/Zeta-Policy} }
 and provide analysis supporting its superior performance compared to $\epsilon$-greedy and some other popular policies. 

A major challenge of deep RL is that it often requires a prohibitively large amount of training time and data to reach a reasonable accuracy, making it inapplicable in real-world settings~\cite{CruzJr2019Pre-trainingLearning}. Leveraging humans to provide demonstrations (known as learning from demonstration (LfD)) in RL has recently gained traction as a possible way of speeding up deep RL~\cite{vinyals2017starcraft,hester2018deep,kurin2017atari}. In  LfD, 
actions demonstrated by the human are considered as the ground truth labels for a given input game/image frame. 
An agent closely simulates the demonstrator's policy at the start, and later on, 
learns to surpass the demonstrator~\cite{CruzJr2019Pre-trainingLearning}. However, LfD holds a distinct challenge, in the sense that it often requires the agent to acquire skills from only a few demonstrations and interactions due to the time and expense of acquiring them~\cite{Calinon2018LearningDemonstration}. Therefore,
LfDs are generally not scalable, especially for high-dimensional problems. 

We propose the technique of pre-training the underlying deep neural networks to speed up training in deep RL. It enables the RL agent to learn better features leading to better performance without changing the policy learning strategies \cite{CruzJr2019Pre-trainingLearning}.
In supervised methods, pre-training helps regularisation and enables faster convergence compared to randomly initialised networks \cite{yu2010roles}. Various studies (e.\,g., \cite{thomas2013deep,liu2014graph}) have explored pre-training in speech recognition and achieved improved results. However, pre-training in deep RL is hardly explored in the area of 
speech \TR{emotion} recognition. In this paper, we present the analysis showing that pre-training can reduce the training time. In our envisioned scenario, the agent might be trained with one corpus but might need to interact with other corpora. To test the performance in those scenarios, we also analyse performance for cross-corpus pre-training.


\section{Related Work and Background}
\RR{Reinforcement learning has not been widely explored for speech emotion recognition. The closest match of our work is EmoRL where the authors use deep RL to determine the best position to split the utterance to send to the emotion recognition model \cite{Lakomkin2018EmoRL:Learning}. The RL agent chooses to decide an ``action'' whether to wait for more audio data or terminate and trigger prediction. Once the terminate action is selected, the agent stops processing the audio stream and starts classifying the emotion. 
The authors aim to achieve a trade-off between accuracy and latency by penalising wrong classifications as well delayed predictions through rewards. In contrast, our focus is on developing a new policy tailored for SER, and apply pre-training to achieve faster learning rate.}

\RR{Another related study has used RL to develop a music recommendation system based on the mood of the listener. The users select their current mood by selecting ``pleasure'' and ``energy'', then, the application selects a song out of its repository. The users can provide feedback on the system recommended audio by answering the question ``Does this audio match the mood you set?'' \cite{Stockholm2009ReinforcementAudio}. Here, the key focus is to learn the mapping of a song to the selected mood, however, in this article, we focus on the automatic determination of the emotion. Yu and Yang introduced an emotion-based target reward function~\cite{Yu2019AnDesign}, which again did not have a focus on SER.} 


\RR{A group of studies used RL for audio enhancement. Shen et al.\ showed that introducing RL for enhancing the audio utterances can reduce the testing phase character error rate by about 7\,\% \TR{of automatic speech recognition with noise} 
\cite{Shen2019ReinforcementRecognition}. 
Fakoor et al.\ also used RL for speech enhancement while considering the noise suppression as a black box and only taking the feedback from the output as the reward. They achieved an increase of 42\,\% of the signal to noise ratio of the output \cite{Fakoor2018ReinforcementQuality}.}

\RR{On the topic of proposing new policies, Lagoudakis and Parr used a modification of approximate policy iteration for pendulum balancing and bicycle riding domains  \cite{Lagoudakis2003ReinforcementClassifiers}. We propose a policy which is tailored for SER. Many studies can be found in the literature using Deep learning to capture emotion and related features from speech signal, however, none of them had a focus on RL~\cite{Han2018ASignal, Latif2018AdversarialRobustness, Rana2019AutomatedFuture}.}

\RR{An important aspect of our study is incorporating pre-training in RL to achieve a faster learning rate in SER. Gabriel et al.\ used human demonstrations of playing Atari games to pre-train a deep RL agent to improve the training time of the RL agent \cite{CruzJr2019Pre-trainingLearning}. Hester et al.\ in their introduction of Deep Q-Learning from Demonstrations, presented prior recorded demonstrations to a deep RL system of playing 42 games -- 41 of them had improved performance and 11 out of them achieved state-of-the-art performance \cite{Hester2017DeepDemonstrations}. 
An evidential advance in performance and accuracy is observed in their results of the case study with Speech Commands Dataset. None of these studies focuses on using pre-training in RL for SER, which is our focus.} 

\subsection{Reinforcement Learning}
RL architecture mainly consist of two major components namely ``Environment'' and ``Agent'' while there are three major signals passing between those two agents as ``current state'', ``reward'' and ``next state''. 
An agent interacts with an unknown environment and observes a state $s_{t}\in S$ at every time step $t \in [0,T]$, where $S$ is the state space and $T$ is the terminal time. The agent selects an action ${a \in A}$, where $A$ is the action space and then, the environment changes to $s_{t+1}$ and the agent receives a reward scalar $r_{t+1}$ which represents a feedback on how good the selected action on the environment is. The agent learns a policy $\pi$ to map most actions $a$ for a given state $s$. The objective of RL is to learn $\pi^*$, an optimal policy which maximises the cumulative reward that can map the most suitable actions to a given state $s$. 


\subsubsection{Q-Learning}
\RR{Before introducing Q-Learning, introduction to a number of key terms is necessary.}

\RR{\noindent{\bf Model Free and Model based Policy:}
The model learns the transition probability from the current state for an action to the next state. Although it is straight forward, model-based algorithms become impractical as the state space and action space grows. In contrary, model-free algorithms rely on trial-and-error to update its knowledge and does not require to store all the combination of states and actions.}

\RR{\noindent{\bf On-Policy learning and Off-policy learning:} These two policies can be described in-terms of Target Policy and Behaviour Policy. Target Policy is the policy, which an agent tries to learn by learning value function. Behavior Policy is used by an agent for action selection or in other words used to interact with the environment.} 

\RR{In On-policy learning, target policy and the behavioural policies are the same, but different in Off-policy learning. Off-policy learning enables continuous exploration resulting in learning an optimal policy, whereas on-policy learning can only offer learning sub-optimal policy.} 


In this paper, we consider widely used Q-Learning, which represents a model-free off-policy learning.
Q-Learning maintains a Q-table which contains Q-values for each state, and action combination. This Q-table is updated each time the agent receives a reward from the environment. A straight forward way to store this information would be in a Table \ref{tab:Q_table}.
\begin{table}[t]
    \fontsize{10}{14}\selectfont
    \centering
    \caption{Fundamental Structure of Q-Table}
    \label{tab:Q_table}
    \begin{tabular}{|c|c|c|}
        \hline
        state & action & $Q(state,action)$ \\ \hline
        ...   & ...    & ...               \\ \hline
        \end{tabular}%
\end{table}
Q-values from the Q-table are used for the action selection process.
A policy takes Q-values as input and outputs the action to be selected by the agent. Some of the policies are Epsilon-Greedy policy \cite{Watkins1989LearningRewards}, Boltzmann Q Policy, Max-Boltzmann Q Policy \cite{Wiering1999ExplorationsLearning}, and Boltzmann-Gumbel exploration Policy \cite{Cesa-Bianchi2017BoltzmannRight}. 

The main disadvantage of Q-Learning is that the state space should be discrete and the Q-table cannot store Q-values for continues state space. Another disadvantage is that the Q-table grows larger with increasing state space which will not be manageable at a certain point. This is known as ``curse of dimensionality'' \cite{Wiering1999ExplorationsLearning} and the complexity of evaluating a policy scales up with $O(n^3)$ when $n$ is the number of states in a problem. Deep Q-Learning offers a solution to these challenges.

\subsubsection{Deep Q-Learning}
A Neural Network can be used to approximate the Q-values based on the state as input. This is more tractable than storing every possible state in a table like Table \ref{tab:Q_table}. 
\begin{equation}
	\label{eq:dqn_Q}
	Q = neural\_network.predict(state)
\end{equation}
When the problems and states are becoming more complex, the Neural Network may need to be ``deep'', meaning a few hidden layers may not suffice to capture all the intricate details of that knowledge, hence the use of Deep Neural Networks (DNNs).
Studies were carried out to incorporate DNNs to Q-Learning by replacing the Q-table with a DNN known as deep Q-Learning,
and the involvement of Deep Learning to Reinforcement Learning is now known as Deep Reinforcement Learning. 

\RR{Deep Q-Learning process uses two neural networks models: the inferring model and the target networks. These networks have the same architecture but different weights. Every $N$ steps,	 the weights from the inferring network are copied to the target network. Using both of these networks leads to more stability in the learning process and helps the algorithm to learn more effectively.}

\subsubsection{Warm-up period}
Since the RL agent learns only by interacting with the environment and the gained reward, the RL agent needs a set of experiences to start training for the experience replay. A parameter ${nb\_warm\_up}$ is used to define the number of warm-up steps to be performed before initiating RL training. 
During this period, actions are selected totally through a random function for a given state, and no Q-value is involved. The state, action, reward, and next state are stored in the memory buffer and are used to sample the experience replay. 

\subsubsection{Q-Learning Policies}
Q-Learning policies are responsible for deciding the best action to be selected based on the Q-values as their input. Exploration and exploitation are used to improve the experience of the experience replay by involving randomness to the action selection. Epsilon-Greedy policy,  Max-Boltzmann Q-Policy, and Linear Annealed wrapper on Epsilon-Greedy policy are some of the Q-Learning policies used today~\cite{Pan2020ReinforcementUpdates, Leibfried2018Model-BasedNetworks}.  

The Epsilon-Greedy policy adopts a greedy algorithm to select an action out from the Q values, The $\epsilon$ value of this policy determines the exploitation and exploration ratio. $1-\epsilon$ is the probability of choosing exploitation on action selection. A random action is selected by a uniform random distribution at exploration and an action with maximum Q-value is selected on exploitation. 
The Linear Annealed wrapper on the Epsilon-Greedy policy is changing the $\epsilon$ value of the Epsilon-Greedy policy at each step. An $\epsilon$ value range and the number of steps are given as parameters. This wrapper linearly changes and updates the $\epsilon$ value of the Epsilon-Greedy policy at each step. 

\TR{The Max-Boltzmann policy also uses $\epsilon$ as a parameter. $\epsilon$ value is considered when determining exploitation and exploration. Exploration in Max-Boltzmann policy is similar as the Epsilon-Greedy policy. At exploitation, instead of selecting the action with maximum Q-value as in Epsilon-Greedy policy; Max-Boltzmann policy randomly selects an action from a distribution which is similar to the Q-values. This introduces more randomness yet, usage of Q-values in to the action selection process.}


\subsection{Scope of RL in Speech Emotion Recognition}
Emotion recognition from speech has gained a fair amount of attention over the past years among machine learning researchers and many studies are being carried out to improve the performance of SER from both feature extraction and emotion classification stages \cite{ElAyadi2011SurveyDatabases}. Hidden Markov Models, Support Vector Machines, Gaussian Mixture Models, and Artificial Neural Networks are some of the classifiers used for SER in the literature \cite{Ververidis2006EmotionalMethods, Schuller2003HiddenRecognition, Cairns1994NonlinearConditions, Lee2002CombiningRecognition}. 
With the tremendous success of DNN architectures, numerous studies have successfully used them and achieved good performances, e.g., \cite{Latif2019DirectSpeech, Han2014SpeechMachine, Latif2018CrossTechnique, Huang2019SpeechSounds}. 


\RR{Using DNN, Supervised and guided unsupervised/self-supervised techniques are being dominantly developed for SER, however, there is still a gap in the literature for dynamically updating SER systems. Although some studies, e.\,g., Learning$++$~\cite{983933} and Bagging$++$~\cite{zhao2010incremental} use incremental learning/Online learning, there is a major difference between Online learning and RL. Online learning is usually used for a constant stream of data, where after once using an example, it is discarded. Whereas, RL constitutes a series of state-action pairs that either draw a positive or a negative reward. If it draws a positive reward, the entire string of actions leading up to that positive reward is reinforced, but if it draws a negative reward, the actions are penalised.}


\begin{figure}[htp]
	\centering
	\includegraphics[width=88mm]{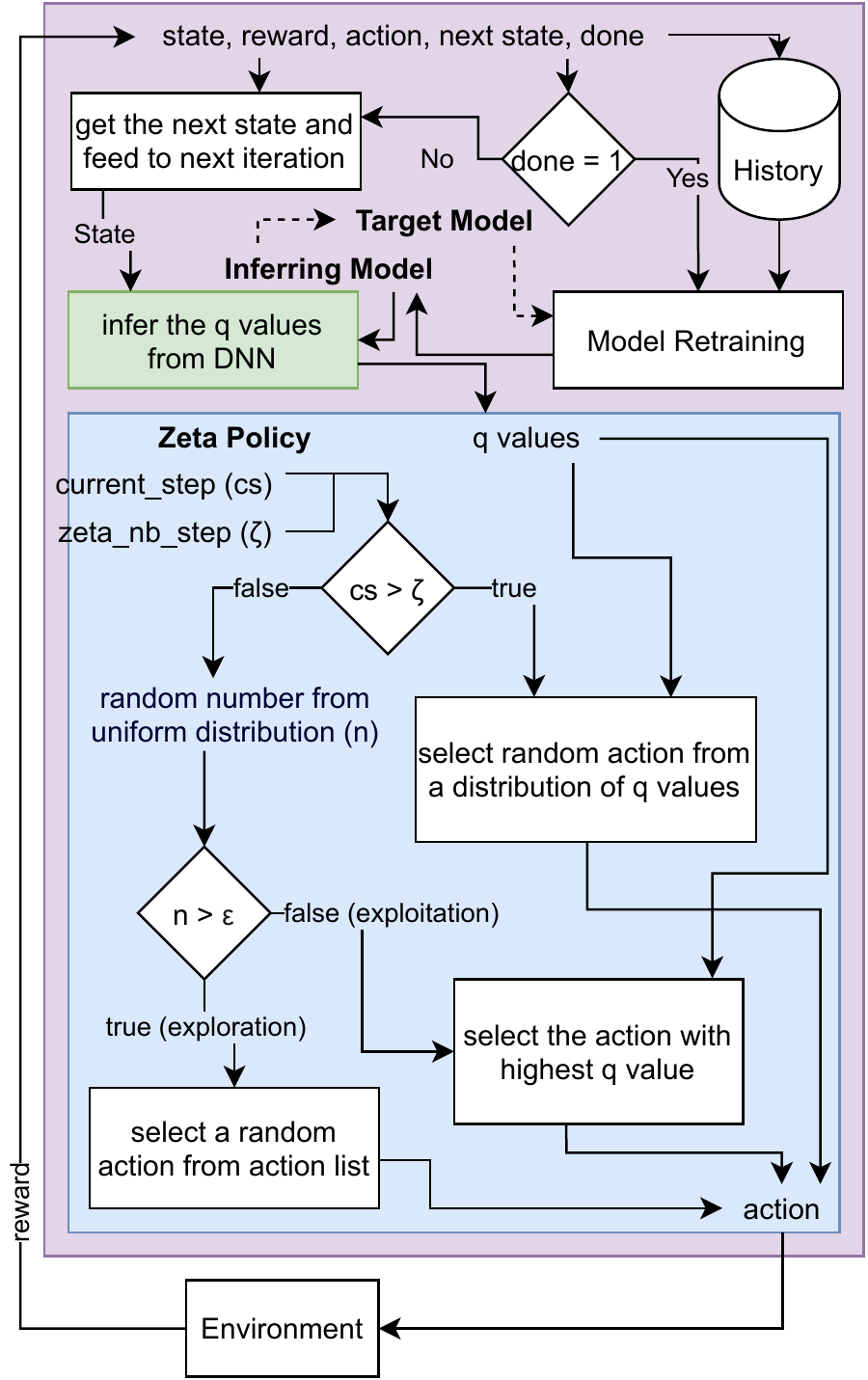}
	\caption{
	Flow of the Zeta Policy and the connection with related components in the RL Architecture.}
	\label{fig:RL_Architecture_Zeta}
\end{figure}

\section{Methodology}

\subsection{Model Process Flow}
\RR{Figure \ref{fig:RL_Architecture_Zeta} shows our proposed architecture and process flow. 
It dominantly focuses on the process flow with the ``Zeta Policy'', yet showing the inter connection between the inferring and the target model and the interaction between the agent and the environment through action and reward. Through out this section we will gradually elaborate on different components.}

\RR{Inferring model is used to approximate the Q-values used in action selection for a given state by the environment. After a reward signal is received by the agent after an action is executed in the environment, attributes (state, reward, action, next state and terminal signal) are stored in the experience memory (History DB in Figure~\ref{fig:RL_Architecture_Zeta}). After every $N_{update}$ number of steps, a batch of samples from the experience memory is used and updates the parameters of Inferring Model. 
	Target Model is used to approximate the $Q_{target}$ values. Usage of $Q_{target}$ value is explained in section~\ref{sec:model_parameter_update}. Parameters of the Target model is updated after $N_{copy}$ number of steps. $N_{update}$ and $N_{copy}$ are hyper-parameters.}

\subsection{Zeta Policy}

A novel RL policy: ``Zeta Policy'' is introduced in this study, which takes inspiration from the Boltzmann Q Policy and the Max-Boltzmann Q Policy. This policy uses the $current\_ step$ ($cs$) (see Figure \ref{fig:RL_Architecture_Zeta}) of the RL training cycle and decides on how the action selection is performed. Figure \ref{fig:RL_Architecture_Zeta} shows how the action selection process is performed by the Zeta Policy with the connection through other components in the RL architecture. The $Zeta\_nb\_step (\zeta)$ is a hyper-parameter to the policy and it routes to the action selection process.  If $cs < \zeta$, the policy follows an exploitation and exploration process, where exploitation selects the action with the highest Q-value and exploration selects a random action from a discrete uniform random distribution to include uniform randomness to the experiences.
\TR{The parameter $\epsilon$ compared with a random number $n$ from an uniform distribution and} used to determine the exploration and exploration route. If $cs > \zeta$, a random value from a distribution similar to the Q-value distribution is picked as the selected action. Experiments were carried out to find the effect of the parameters $\zeta$ and $\epsilon$ on the performance of RL. 

In the SER context, a state $s$ is defined as an utterance in the dataset and an action $a$ is the classified label (emotion) to the state $s$. A reward $r$ is the reward returned by the environment after comparing the ground truth with the action $a$. If an action (classified emotion) and ground truth are similar, i.\,e., the agent has inferred the correct label to the given state, the \HL{reward resembles $1$ and else the reward is $-1$. }
The reward is accumulated throughout the episode, and the mean reward is calculated. The higher the mean reward, the better the performance of the RL agent. The standard deviation of the reward is also calculated, since this value interprets how robust the RL predictions are. 

\subsection{Pre-training for improved performance}
Pre-training allows the RL agent's inferring DNN to optimise its parameters and learning the features required for a similar problem. To use a pre-trained DNN in RL, we replace the softmax layer with a $Q$-value output layer. As the inferring model is optimised with learnt features, there is no necessity of a warm-up period to collect experience replay. In fact, extended training time is a key shortcoming of RL \cite{CruzJr2019Pre-trainingLearning}.
One key contribution of the present paper is to use pre-training to reduce the training time by reducing the warm-up period.




\begin{figure}[!ht]
	\centering
	\includegraphics[trim=0cm 0cm 0cm 0cm,clip=true,width=.5\textwidth]{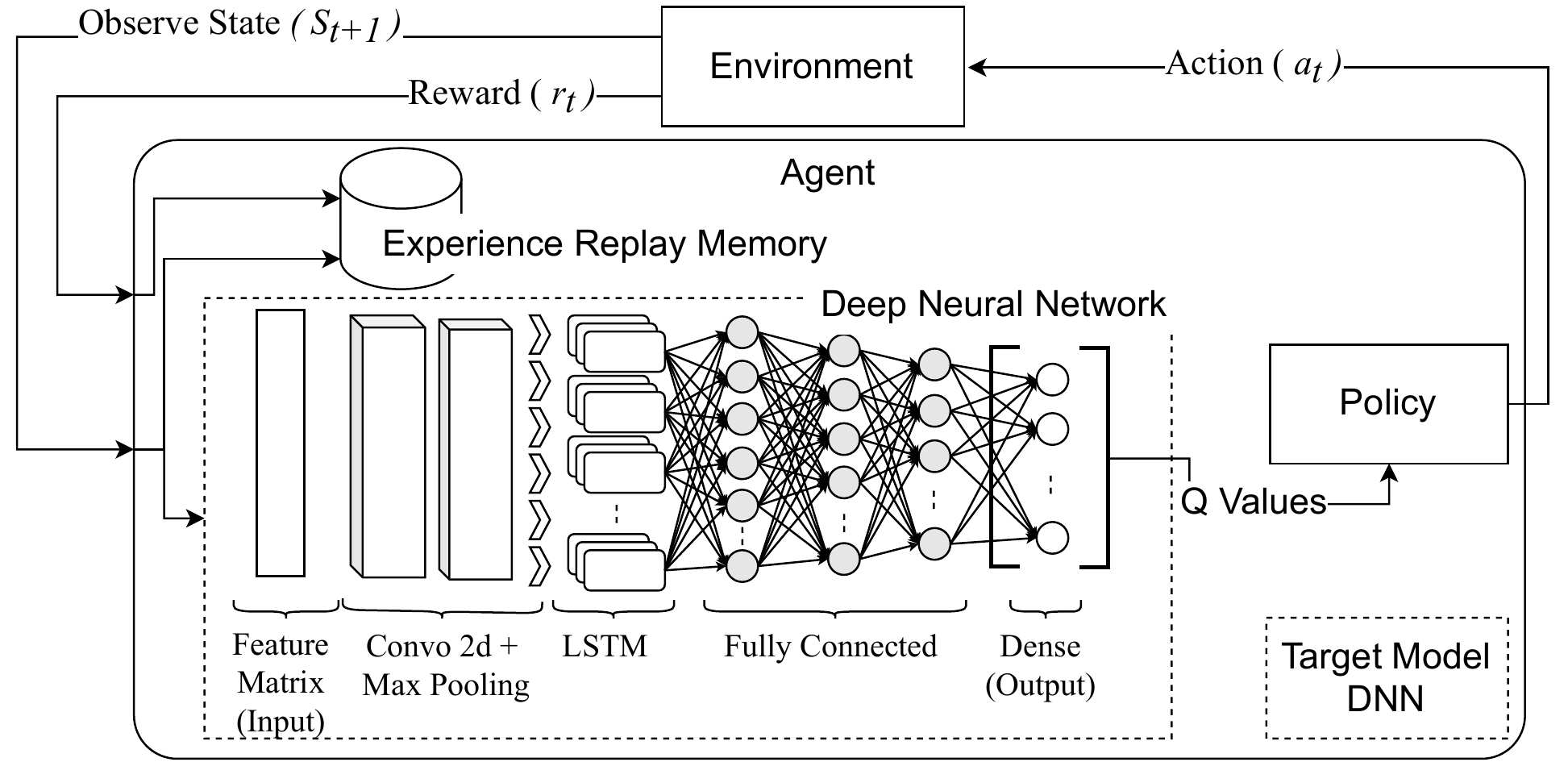}
	\caption{Deep Reinforcement Learning Architecture}
	\label{fig:deep_rl_architecture}
	\vspace*{-5mm}
\end{figure}


\subsection{Experimental Setup}
In this section, we explain the experimental setup including feature extraction and model configuration. Figure~\ref{fig:deep_rl_architecture} shows the experimental setup architecture of the Environment, Agent, Deep Neural Network and Policy used in this study.

\subsubsection{Dataset}
\begin{table}[t]
    \fontsize{10}{14}\selectfont
	\centering
	\caption{Distribution of utterances in the two considered datasets by emotion}
	\label{tab:dataset_discription}
	\begin{tabular}{l|c|c}
		\hline
		\textbf{Emotion} & \multicolumn{1}{l|}{\textbf{IEMOCAP}} & \multicolumn{1}{l}{\textbf{SAVEE}} \\ \hline
		Happy   & 895  & 60  \\
		Sad     & 1200 & 60  \\
		Angry   & 1200 & 60  \\
		Neutral & 1200 & 120 \\ \hline
	\end{tabular}%
\vspace*{-5mm}
\end{table}

This study uses two popular datasets in SER: IEMOCAP \cite{Busso2008IEMOCAP:Database} and  SAVEE \cite{Haq2008Audio-visualClassification}.
The IEMOCAP dataset features five sessions; each session includes speech segments from two speakers and is labelled with nine emotional categories. However, we use happiness, sadness, anger, and neutral for consistency with the literature. 
The dataset was collected from ten speakers (five male and five female). We took a maximum of 1\,200 segments from each emotion for equal distribution of emotions within the dataset. 

Note that the SAVEE dataset is relatively smaller compared to IEMOCAP. It is collected form 4 male speakers and has 8 labels for emotions which we filtered out keeping happiness, sadness, anger, and neutral segments for alignment with IEMOCAP and the literature. 

Table \ref{tab:dataset_discription} shows the utterances'  distribution of the two datasets with 30\,\% of each dataset being used as a subset for pre-training, and the rest being used for the RL execution. 

\TR{20\% of the RL execution data subset is used in the testing phase. RL Testing phase executes the whole pipeline of RL algorithm but does not update the model parameters.
	RL which is a different paradigm of Machine Learning than Supervised Learning does not need to have a testing dataset as it contentiously interacting with an environment. But since this specific study is related to a classification problem, we included a testing phase by proving a dataset which the RL agent has not seen before. }


We remove the silent sections of the audio segments and process only the initial two seconds of the audio. For segments less than two seconds in length, we use zero padding. 
The reason is when using the two seconds for segment length, there will be more zero-padded segments if the segment length is increased and it becomes an identifiable feature within the dataset in the model training.


\subsubsection{Feature Extraction}
We use Mel Frequency Cepstral Coefficients (MFCC) to represent the speech utterances. MFCC are widely used in speech audio analysis \cite{Latif2019DirectSpeech, Davis1980ComparisonSentences}. We extract 40 MFCCs from the Mel-spectrograms with a frame length of 2,048 and a hop length of 512 using Librosa \cite{McFee2015Librosa:Python}. 
\begin{figure*}[!ht]
	\centering
	\includegraphics[trim=0cm 0cm 0cm 0cm,clip=true,width=\textwidth]{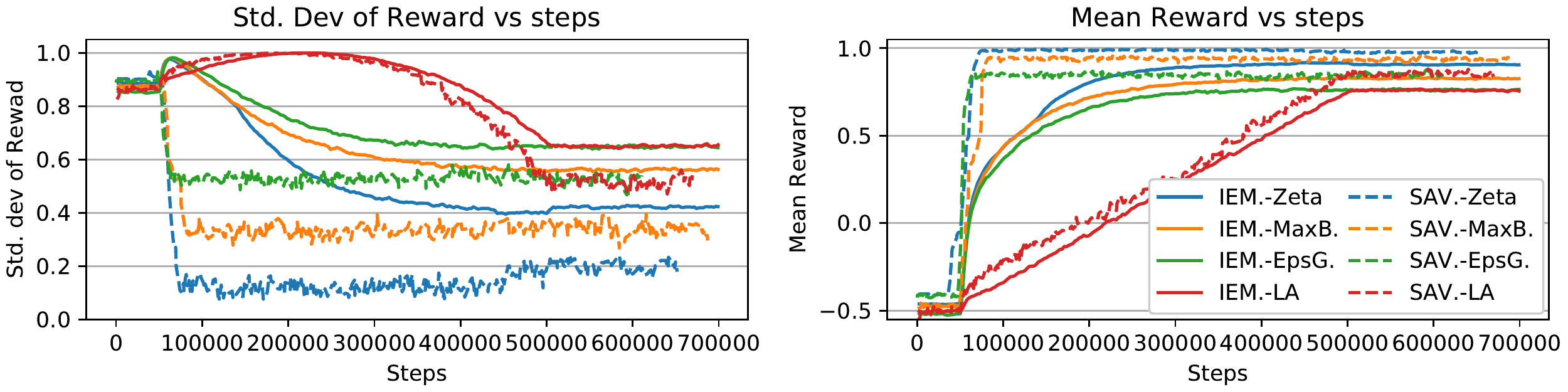}
	\caption{Comparison across the policies Zeta policy $(Zeta)$, Epsilon-Greedy policy $(EpsG.)$, Max-Boltzmann Q-policy $(MaxB.)$, and Linear Annealed wrapper on Epsilon-Greedy policy $(LA)$ for the two datasets IEMOCAP $(IEM)$ and SAVEE $(SAV)$ }
	\label{fig:comparision_of_policies}
\end{figure*}

\begin{figure*}[!ht]
	\centering
	\includegraphics[trim=0cm 0cm 0cm 0cm,clip=true,width=\textwidth]{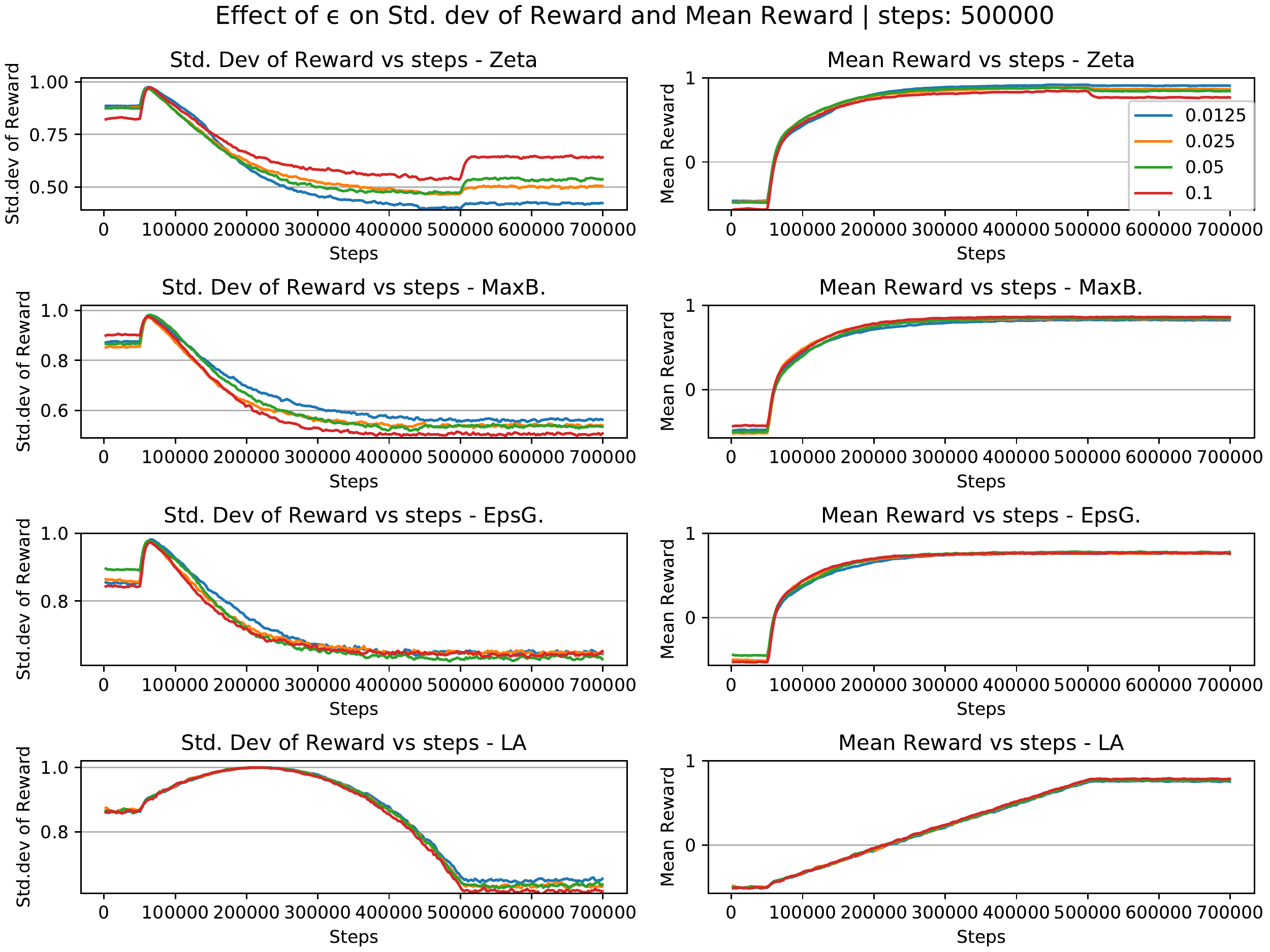}
	\caption{Comparison of mean reward and standard deviation of reward with policies by changing the $\epsilon$ value of the Zeta policy $(Zeta)$, the Epsilon-Greedy policy $(EpsG.)$, the Max-Boltzmann Q-policy $(MaxB.)$, and Linear Annealed wrapper on the Epsilon-Greedy policy $(LA)$.}
	\label{fig:comparision_changing_eps}
	\vspace*{-5mm}
\end{figure*} 
\begin{figure*}[!ht]
	\centering
	\includegraphics[trim=0cm 0cm 0cm 0cm,clip=true,width=\textwidth]{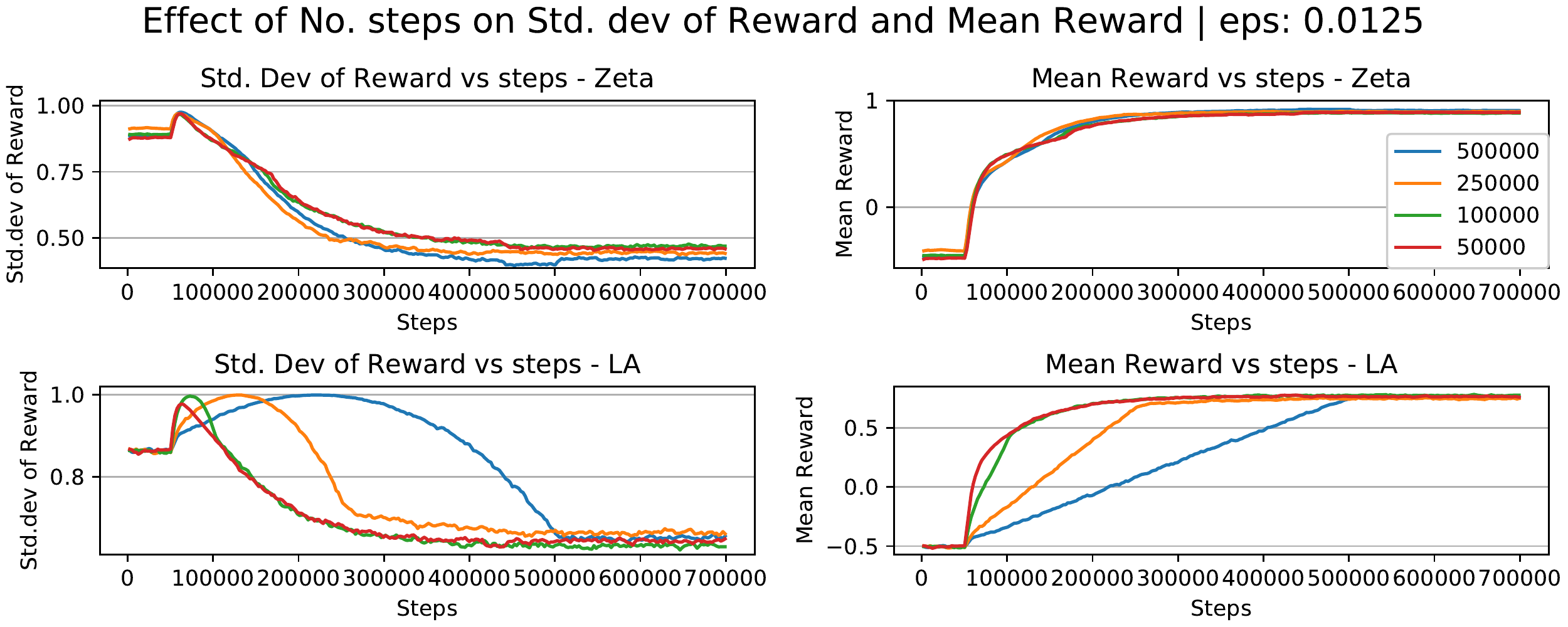}
	\caption{Comparison of mean reward and standard deviation of reward with policies by changing the number of steps of the Zeta Policy $(Zeta)$ and the Linear Annealed wrapper on the Epsilon-Greedy policy $(LA)$.}
	\label{fig:comparision_changing_no_steps}
	\vspace*{-5mm}
\end{figure*}

\subsubsection{Model Recipe} \label{subsection:model_recipe}



\TR{A supervised learning approach is followed to identify the best suited DNN model as the inferring model. Even-though the last layer of supervised learning models output a probability vector, RL also learn the representaions in the hidden layers with similar mechanism. DNN architecture of the supervised learning model is similar to the DNN architecture of the inferring model except the output later. Activation function of the output layer of supervised later is Softmax whereas it is Linear Activation in RL. 
	Different architectures containing Convolutional Neural Networks (CNNs),  Long-Short Term Memory (LSTM) Recurrent Neural Networks (RNNs) and Dense Layers were evaluated with similar dataset and DNN model architecture with highest testing accuracy was selected as the Inferring DNN architecture. }

We use the popular Deep Learning API Keras~\cite{Chollet2015Keras} with TensorFlow~\cite{Abadi2015TensorFlow:Systems} as the backend 
for modelling and training purposes in this study. We model the RL agent's DNN with a combination of CNNs and  LSTM. The use of a CNN-LSTM combined model is motivated by the ability to learn temporal and frequency components in the speech signal \cite{Latif2020DeepTrends}. We stack the LSTM on a CNN layer and pass it on to a set of fully connected layers to learn  discriminative features \cite{Latif2019DirectSpeech}. 

As shown in Deep Neural Network component in Figure~\ref{fig:deep_rl_architecture}, the initial layers of the inferring DNN model are two 2D-convolutional layers with filter sizes 5 and 3, respectively. Batch Normalisation is applied after the first 2D convolutional layer. The output from the second 2D convolutional layer is passed on to an LSTM layer of 16 cells, then to a fully connected layer of 265 units. A dropout layer of rate 0.3 is applied \TR{before the dense layer which outputs the Q-values.} \HL{Number of outputs in the dense layer is equal to the the number of actions which is number of classes of classification.} \TR{Activation function of the last layer is kept as linear function since the output Q-values should not be normalised.} 
The input shape of the model is $40 \times 87$, where 40 is the number of MFCCs and 87 is the number of frames in the MFCC spectrum. An Adam optimiser with a learning rate 0.00025 is used when compiling the model.


\TR{The output layer of the neural network model returns an array of Q-values corresponding to each action for a given state as input. All the Q learning policies used in the paper (Zeta, Epsilon-Greedy, Max-Boltzmann Q Policy, and Linear Annealed wrapper on Epsilon-Greedy) then use these output Q-values for corresponding action selection.} \TR{We use the number of steps on executing Inferring model update ($N_{update}$) equal to 4 and number of steps on executing parameter copy from Inferring model to Target model ($N_{copy}$) equal to 10\,000. }

\TR{Supervised learning is used to pre-train the inferring model. A DNN which is similar architecture of the inferring model, but with Softmax activation function at output layer is trained with pre-training data subset before starting the RL execution. The model is trained for 64 epochs with a batch size 128.
	Once the pre-training is completed parameters of the pre-trained DNN model is copied to the inferring and target models in the RL Agent.} 

\subsubsection{Model parameter update}
\label{sec:model_parameter_update}
\TR{Mathematical formulation of Q-Learning is based on the popular Bellman's Equation for State-Value function (Eq~\ref{eq:bellmans_value_function}). In~\eqref{eq:bellmans_value_function}, $v(s)$ is the value at state $s$, $R_t$ is the reward at time $t$ and $\gamma$ is the discount factor. }
\begin{equation}
	\label{eq:bellmans_value_function}
	v(s)  = E[R_{t+1} +\gamma v(s_{t+1}) | S_t=s]
\end{equation}
\TR{Equation~\eqref{eq:bellmans_value_function} can be re-written to obtain the Bellman's State-Action Value function known as Q-function as follows; }
\begin{equation}
	\label{eq:bellmans_state_action_value_function}
	q_\pi(s,a) = E_\pi[R_{t+1} + \gamma q_\pi(S_{t+1},A_{t+1}) | S_t=s, A_t=a]
\end{equation}
\TR{Here $q_\pi(s,a)$ is the Q-value of the state $s$ following action $a$ under the policy $\pi$.}

\TR{We use a DNN to approximate the Q-values and $q_\pi(s,a)$ is considered as the target Q-value giving the loss function~\eqref{eq:loss_function}}
\begin{equation}
	\label{eq:loss_function}
	L = \Sigma{(Q_{target} - Q)^2}
\end{equation}
\begin{equation}
	\label{eq:q_target}
	Q_{target} = R(s_{t+1},a_{t+1}) + \gamma Q(s_{t+1},a_{t+1}) 
\end{equation}
\TR{Combining Equations~\eqref{eq:loss_function} and~\eqref{eq:q_target}, updated loss function can be written as;}
\begin{equation}
	\label{eq:final_loss_function}
	L = \Sigma{(R(s_{t+1},a_{t+1}) + \gamma Q(s_{t+1},a_{t+1})  - Q(s_t,a_t) )^2}
\end{equation}
\TR{Minimising the Loss value $L$ is the optimisation problem solved in the training phase. }

\TR{$Q_{target}$ is obtained by inferring the state $s$ from the target network via the function~\eqref{eq:dqn_Q}. The output layer of the DNN model is a Dense layer with Linear activation function. Adopting the feed-forward pass in deep neural networks to the last layer $l$, Q value for an action $a$ can be obtained by equation~\eqref{eq:q_output},}
\begin{equation}
	\label{eq:q_output}
	Q_a = W^lx^{l-1} + b^l
\end{equation}
\TR{where $x^{l-1}$ is the input for the layer $l$. Backpropergation pass in parameter optimising updates the $W^l$ and $b^l$ values in each later $l$ to minimise the Loss value $L$.}
\section{Evaluation}
    
Experiments were carried out to evaluate the efficiency of the newly proposed policy ``Zeta-Policy'' and the improvement that can be gained by introducing the pre-training aspect to the RL paradigm. 

\subsection{Comparison of Policies}
The key contribution presented in this article is the Zeta Policy. Therefore, we first compare its performance with that of three commonly used policies: the Epsilon-Greedy policy (EpsG.), the Max-Boltzmann Q Policy (MaxB.), and Linear Annealed wrapper on Epsilon-Greedy policy \TR{(LA)}. 
Figure \ref{fig:comparision_of_policies} presents the results of these comparisons. Standard deviation and mean reward are plotted against the step number. 


We notice from Figure \ref{fig:comparision_of_policies} that the  Zeta policy has a lower standard deviation of the reward, which suggests that the robustness of the selected actions (i.\,e., classified labels/emotions) of the Zeta policy is higher than that of the compared policies. 

The Zeta policy outperforms the other compared policies with higher mean reward. The mean reward of the Zeta policy converges to a value around 0.78 for the IEMOCAP dataset and 0.7 for the MSP-IMPROV dataset. These values are higher than that of other policies compared, which means that the RL Agent selects the actions more correctly than other policies. Since the inferring DNN model of the RL Agent is the same for all experiments, we can infer that the Zeta policy has played a major role in this out-performance. 
	 

\begin{figure*}[!ht]
	\centering
	\includegraphics[trim=0cm 0cm 0cm 0cm,clip=true,width=\textwidth,height=4.5cm]{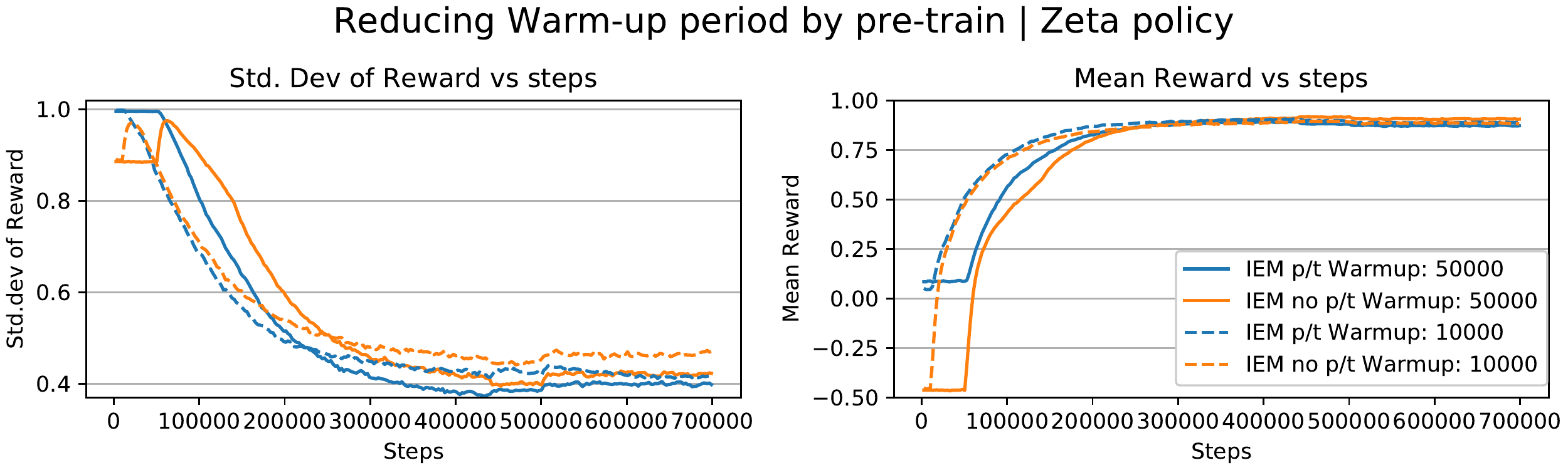}
	\caption{Effect of the warm-up period and pre-training (p/t) on the performance.}
	\label{fig:reducing_wamup_by_pretraining}
\end{figure*}
\begin{figure*}[!ht]
 	\centering
 	\includegraphics[trim=0cm 0cm 0cm 0cm,clip=true,width=\textwidth]{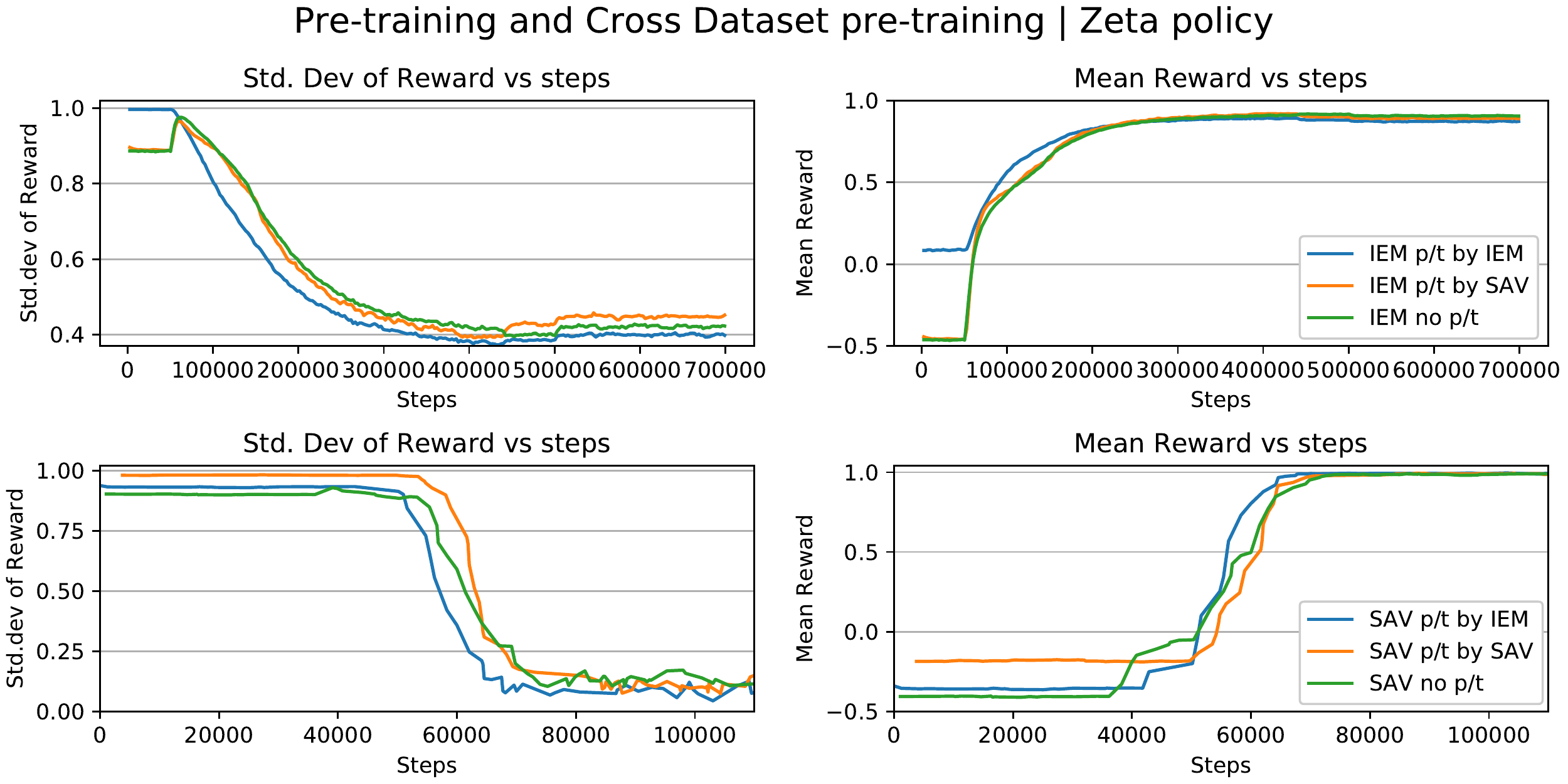}
 	\caption{Performance impact on pre-training (p/t) and cross dataset pre-training by the IEMOCAP $(IEM)$ dataset and the SAVEE $(SAV)$ dataset }
 	\label{fig:cross_dataset_pre_training}
 	\vspace*{-5mm}
\end{figure*}

\subsubsection{Impact of the $\epsilon$ value}

All the compared policies --- Zeta Policy, Epsilon-Greedy policy, Max-Boltzmann $Q$ policy, and the Linear Annealed wrapper on Epsilon-Greedy policy use the parameter $\epsilon$ in their action selection process. $\epsilon$ is used to decide the exploration and exploitation within the policy. $1-\epsilon$ is the probability of selecting exploitation out of exploration and exploitation. Hence, $\epsilon$ ranges between 0 and 1. 

Several experiments were carried out to find the range of $\epsilon$ with the values  $0.1$,  $0.05$, $0.025$, and $0.0125$. Figure \ref{fig:comparision_changing_eps} shows the effect of changing the $\epsilon$ value on the standard deviation of the reward and mean reward for all policies. 

The Zeta policy shows a noticeable change in the standard deviation of the reward and mean reward for $\epsilon = 0.0125$.  The reason being that the Zeta policy performs well in the lower $\epsilon$ scenarios is: when $cs < \zeta$, the Zeta policy picks actions with the exploration and exploitation strategy and $\epsilon$ is the probability of exploration. A random action from a uniform random distribution is picked in the exploration strategy. A lower $\epsilon$ means lower randomness in the period $cs < \zeta$ which has a higher probability of selecting an action based on the $Q$-values which leads the RL algorithms to correct the false predictions. 
	
\subsubsection{Impact of the number of steps}
The Zeta policy 
uses the parameter $Zeta\_nb\_step (\zeta)$ to determine the route and the Linear Annealing uses the parameter $nb\_step$ to determine the gradient of the $\epsilon$ value used in the Epsilon-Greedy component. Experiments were defined to examine the behavior of the performance of the RL agent by changing the above parameters to the values $500\,000$, $250\,000$, $100\,000$ and $50\,000$. Figure \ref{fig:comparision_changing_no_steps} was drawn with the output from experiments. Looking at the curve of the standard deviation of Zeta policy, the robustness of the RL agent has increased with the increase of the number of steps. The graph shows that the most robust curve is observed for the step size $500\,000$. 

\subsection{Pre-training for the reduced warm-up period}
The warm-up period is executed in the RL algorithm to collect experiences for the RL agent to sample the experience replay. But with the pre-training, inferring DNN, the RL agent is already trained to learn the features. This leads the RL agent to produce better $Q$ values than an RL agent inferring a DNN based on randomly initialised parameters. An experiment was executed to identify the possibility of reducing the warm-up period after pre-training and yet keep the RL agent performance unchanged shown in Figure \ref{fig:reducing_wamup_by_pretraining} that features the generated results. Observing both standard deviation of the reward and mean reward in Figure \ref{fig:reducing_wamup_by_pretraining}, pre-training has improved the robustness and performance of the prediction. The time taken to achieve the highest performance has reduced since the warm-up period is reduced. This makes the RL training time lower and time taken for optimising the RL agent as well. 

\TR{Speed-up of the training period by pre-training was calculated by considering the number of steps needed to reach the mean reward value of 0.6. Mean number of steps taken to reach mean reward of 0.6 without pre-training was 77126 whilst with pre-training (warm-up 10000) was 48623. The speed-up of training period by pre-training and reducing warm-up steps was 1.63x.} 

\subsection{Cross-Dataset pre-training}
In our envisioned scenario an agent, although pre-trained with one corpus, is expected to be robust to other corpus/dialects. In order to experiment the behaviour of cross dataset pre-training on RL, we pre-trained the RL Agent with SAVEE pre-train data subset for the IEMOCAP RL agent and plotted the reward curve and standard deviation of the reward curve in Figure \ref{fig:cross_dataset_pre_training}. The graph shows that the pre-training has always improved the performance of the RL agent and cross dataset pre-training has not degraded the performance drastically. This practice can be used in real-world applications with RL implementations, where there is a lower number of training data available. The RL agent can be pre-trained with a dataset which is aligned with the problem and deployed for act with the real environment. 

\subsection{Accuracy of the predictions}
\TR{Accuracy of a machine learning model is a popular attribute that uses to benchmark against the other parallel models. Since RL is a method of dynamic programming, it does not comprise of accuracy attribute. But, as this specific study is focused on a classification problem, we calculated the accuracy of the RL agent with the logs of the environment. Equation~\ref{eq:accuracy} is used to calculate the accuracy value of an episode. 
	Accuracy value of the testing phase after RL execution of 700000 steps was calculated and tabulated in the Table~\ref{tab:accuracies}.}
\begin{equation}
	\label{eq:accuracy}
	\text{accuracy} = \frac{\text{No. correct inferences}}{\text{No. of utterances}} \times 100\%
\end{equation}


\begin{table}[t]
\fontsize{10}{14}\selectfont
\centering
\caption{Testing accuracy values of the two datasets IEMOCAP and SAVEE under each policy after 700000 steps of RL training.}
\label{tab:accuracies}
\begin{tabular}{l|l|l}
\multicolumn{1}{c|}{\textbf{Policy\textbackslash{}Dataset}} & \multicolumn{1}{c|}{\textbf{IEMOCAP}} & \multicolumn{1}{c}{\textbf{SAVEE}} \\ \hline
\textbf{Zeta}   & \textbf{54.29 $\pm$ 2.50} & \textbf{68.90 $\pm$ 0.61} \\
Max-Boltzmann   & 51.92$\pm$ 1.40          & 67.90 $\pm$ 0.40         \\
Epsilon-Greedy  & 51.45 $\pm$ 0.74          & 58.94 $\pm$ 2.85          \\
Linear Annealed & 51.72 $\pm$ 0.75          & 62.20 $\pm$ 6.10          \\ \hline
\end{tabular}%
\vspace*{-5mm}.
\end{table}

\TR{Studying the Table~\ref{tab:accuracies}, it is observed that the Zeta policy outperforms the other compared policies in both datasets. Also, these results can be compared with the results of supervised learning methods even though they are diverse machine learning paradigms.}

\section{Conclusion}
This study was carried out to discover the feasibility of using a novel reinforcement learning policy named as Zeta policy for speech emotion recognition problems. Pre-training the RL agent was also studied to reduce the training time and minimise the warm-up period. The evaluated results show that the proposed Zeta policy performs better than the existing policies. We also provided an analysis of the relevant parameters $epsilon$ and the number of steps, which shows the operating range of these two parameters. The results also confirm that pre-training can reduce the training time to reach maximum performance by reducing the warm-up period. We show that the proposed Zeta Policy with pre-training is robust to a cross-corpus scenario. In the future, one should study a cross-language scenario and explore the feasibility of using the novel Zeta policy with other RL algorithms. 

%
\bibliographystyle{IEEEtran}
\bibliography{references_mandeley, other_ref}

\begin{thebibliography}{10}
\providecommand{\url}[1]{#1}
\csname url@samestyle\endcsname
\providecommand{\newblock}{\relax}
\providecommand{\bibinfo}[2]{#2}
\providecommand{\BIBentrySTDinterwordspacing}{\spaceskip=0pt\relax}
\providecommand{\BIBentryALTinterwordstretchfactor}{4}
\providecommand{\BIBentryALTinterwordspacing}{\spaceskip=\fontdimen2\font plus
\BIBentryALTinterwordstretchfactor\fontdimen3\font minus
  \fontdimen4\font\relax}
\providecommand{\BIBforeignlanguage}[2]{{%
\expandafter\ifx\csname l@#1\endcsname\relax
\typeout{** WARNING: IEEEtran.bst: No hyphenation pattern has been}%
\typeout{** loaded for the language `#1'. Using the pattern for}%
\typeout{** the default language instead.}%
\else
\language=\csname l@#1\endcsname
\fi
#2}}
\providecommand{\BIBdecl}{\relax}
\BIBdecl

\bibitem{Silver2016MasteringSearch}
\BIBentryALTinterwordspacing
D.~Silver, A.~Huang, C.~J. Maddison, A.~Guez, L.~Sifre, G.~van~den Driessche,
  J.~Schrittwieser, I.~Antonoglou, V.~Panneershelvam, M.~Lanctot, S.~Dieleman,
  D.~Grewe, J.~Nham, N.~Kalchbrenner, I.~Sutskever, T.~Lillicrap, M.~Leach,
  K.~Kavukcuoglu, T.~Graepel, and D.~Hassabis, ``{Mastering the game of Go with
  deep neural networks and tree search},'' \emph{Nature}, vol. 529, no. 7587,
  pp. 484--489, 2016. [Online]. Available:
  \url{https://doi.org/10.1038/nature16961}
\BIBentrySTDinterwordspacing

\bibitem{Vinyals2019GrandmasterLearning}
\BIBentryALTinterwordspacing
O.~Vinyals, I.~Babuschkin, W.~M. Czarnecki, M.~Mathieu, A.~Dudzik, J.~Chung,
  D.~H. Choi, R.~Powell, T.~Ewalds, P.~Georgiev, J.~Oh, D.~Horgan, M.~Kroiss,
  I.~Danihelka, A.~Huang, L.~Sifre, T.~Cai, J.~P. Agapiou, M.~Jaderberg, A.~S.
  Vezhnevets, R.~Leblond, T.~Pohlen, V.~Dalibard, D.~Budden, Y.~Sulsky,
  J.~Molloy, T.~L. Paine, C.~Gulcehre, Z.~Wang, T.~Pfaff, Y.~Wu, R.~Ring,
  D.~Yogatama, D.~W{\"{u}}nsch, K.~McKinney, O.~Smith, T.~Schaul, T.~Lillicrap,
  K.~Kavukcuoglu, D.~Hassabis, C.~Apps, and D.~Silver,
  ``\BIBforeignlanguage{en}{{Grandmaster level in StarCraft II using
  multi-agent reinforcement learning}},''
  \emph{\BIBforeignlanguage{en}{Nature}}, vol. 575, no. 7782, pp. 350--354,
  2019. [Online]. Available:
  \url{https://www.nature.com/articles/s41586-019-1724-z}
\BIBentrySTDinterwordspacing

\bibitem{Shen2019ReinforcementRecognition}
Y.~Shen, C.~Huang, S.~Wang, Y.~Tsao, H.~Wang, and T.~Chi, ``{Reinforcement
  Learning Based Speech Enhancement for Robust Speech Recognition},'' in
  \emph{ICASSP 2019 - 2019 IEEE International Conference on Acoustics, Speech
  and Signal Processing (ICASSP)}, 2019, pp. 6750--6754.

\bibitem{Fakoor2018ReinforcementQuality}
\BIBentryALTinterwordspacing
R.~Fakoor, X.~He, I.~Tashev, and S.~Zarar, ``{Reinforcement Learning To Adapt
  Speech Enhancement to Instantaneous Input Signal Quality},''
  \emph{arXiv:1711.10791 [cs]}, 2018. [Online]. Available:
  \url{http://arxiv.org/abs/1711.10791}
\BIBentrySTDinterwordspacing

\bibitem{Chung2020Semi-supervisedLearning}
H.~Chung, H.~B. Jeon, and J.~G. Park, ``{Semi-supervised Training for
  Sequence-to-Sequence Speech Recognition Using Reinforcement Learning},'' in
  \emph{Proceedings of the International Joint Conference on Neural
  Networks}.\hskip 1em plus 0.5em minus 0.4em\relax Institute of Electrical and
  Electronics Engineers Inc., 7 2020.

\bibitem{Singh1999ReinforcementSystems.}
S.~P. Singh, M.~J. Kearns, D.~J. Litman, and M.~A. Walker, ``{Reinforcement
  learning for spoken dialogue systems.}'' in \emph{Nips}, 1999, pp. 956--962.

\bibitem{Paek2006ReinforcementDeployment}
T.~Paek, ``{Reinforcement learning for spoken dialogue systems: Comparing
  strengths and weaknesses for practical deployment},'' in \emph{Proc.
  Dialog-on-Dialog Workshop, Interspeech}, 2006.

\bibitem{Lakomkin2018EmoRL:Learning}
\BIBentryALTinterwordspacing
E.~Lakomkin, M.~A. Zamani, C.~Weber, S.~Magg, and S.~Wermter, ``{EmoRL:
  Continuous Acoustic Emotion Classification using Deep Reinforcement
  Learning},'' \emph{Proceedings - IEEE International Conference on Robotics
  and Automation}, pp. 4445--4450, 4 2018. [Online]. Available:
  \url{http://arxiv.org/abs/1804.04053}
\BIBentrySTDinterwordspacing

\bibitem{Fan2019AQ-Learning}
\BIBentryALTinterwordspacing
J.~Fan, Z.~Wang, Y.~Xie, and Z.~Yang, ``{A Theoretical Analysis of Deep
  Q-Learning},'' \emph{arXiv}, 1 2019. [Online]. Available:
  \url{http://arxiv.org/abs/1901.00137}
\BIBentrySTDinterwordspacing

\bibitem{CruzJr2019Pre-trainingLearning}
\BIBentryALTinterwordspacing
G.~V. d.~l. Cruz~Jr, Y.~Du, and M.~E. Taylor, ``{Pre-training Neural Networks
  with Human Demonstrations for Deep Reinforcement Learning},'' in
  \emph{Adaptive Learning Agents (ALA)}, 2019. [Online]. Available:
  \url{http://arxiv.org/abs/1709.04083}
\BIBentrySTDinterwordspacing

\bibitem{vinyals2017starcraft}
O.~Vinyals, T.~Ewalds, S.~Bartunov, P.~Georgiev, A.~S. Vezhnevets, M.~Yeo,
  A.~Makhzani, H.~K{\"u}ttler, J.~Agapiou, J.~Schrittwieser \emph{et~al.},
  ``Starcraft ii: A new challenge for reinforcement learning,'' \emph{arXiv},
  vol. 2017, no. 1708.04782, 2017.

\bibitem{hester2018deep}
T.~Hester, M.~Vecerik, O.~Pietquin, M.~Lanctot, T.~Schaul, B.~Piot, D.~Horgan,
  J.~Quan, A.~Sendonaris, I.~Osband \emph{et~al.}, ``Deep q-learning from
  demonstrations,'' in \emph{Proceedings AAAI}, 2018.

\bibitem{kurin2017atari}
V.~Kurin, S.~Nowozin, K.~Hofmann, L.~Beyer, and B.~Leibe, ``The atari grand
  challenge dataset,'' \emph{arXiv}, no. 1705.10998, 2017.

\bibitem{Calinon2018LearningDemonstration}
S.~Calinon, ``{Learning from demonstration (programming by demonstration)},''
  \emph{Encyclopedia of Robotics}, pp. 1--8, 2018.

\bibitem{yu2010roles}
D.~Yu, L.~Deng, and G.~Dahl, ``Roles of pre-training and fine-tuning in
  context-dependent dbn-hmms for real-world speech recognition,'' in
  \emph{Proc. NIPS Workshop on Deep Learning and Unsupervised Feature
  Learning}, 2010.

\bibitem{thomas2013deep}
S.~Thomas, M.~L. Seltzer, K.~Church, and H.~Hermansky, ``Deep neural network
  features and semi-supervised training for low resource speech recognition,''
  in \emph{2013 IEEE international conference on acoustics, speech and signal
  processing}.\hskip 1em plus 0.5em minus 0.4em\relax IEEE, 2013, pp.
  6704--6708.

\bibitem{liu2014graph}
Y.~Liu and K.~Kirchhoff, ``Graph-based semi-supervised acoustic modeling in
  dnn-based speech recognition,'' in \emph{2014 IEEE Spoken Language Technology
  Workshop (SLT)}.\hskip 1em plus 0.5em minus 0.4em\relax IEEE, 2014, pp.
  177--182.

\bibitem{Stockholm2009ReinforcementAudio}
J.~Stockholm and P.~Pasquier, ``{Reinforcement Learning of Listener Response
  for Mood Classification of Audio},'' in \emph{2009 International Conference
  on Computational Science and Engineering}, vol.~4, 2009, pp. 849--853.

\bibitem{Yu2019AnDesign}
H.~Yu and P.~Yang, ``{An Emotion-Based Approach to Reinforcement Learning
  Reward Design},'' in \emph{2019 IEEE 16th International Conference on
  Networking, Sensing and Control (ICNSC)}, 2019, pp. 346--351.

\bibitem{Lagoudakis2003ReinforcementClassifiers}
M.~G. Lagoudakis and R.~Parr, ``{Reinforcement learning as classification:
  leveraging modern classifiers},'' ser. ICML'03.\hskip 1em plus 0.5em minus
  0.4em\relax AAAI Press, 2003, pp. 424--431.

\bibitem{Han2018ASignal}
\BIBentryALTinterwordspacing
H.~Han, K.~Byun, and H.~G. Kang, ``{A deep learning-based stress detection
  algorithm with speech signal},'' in \emph{AVSU 2018 - Proceedings of the 2018
  Workshop on Audio-Visual Scene Understanding for Immersive Multimedia,
  Co-located with MM 2018}.\hskip 1em plus 0.5em minus 0.4em\relax New York,
  NY, USA: Association for Computing Machinery, Inc, 10 2018, pp. 11--15.
  [Online]. Available: \url{https://dl.acm.org/doi/10.1145/3264869.3264875}
\BIBentrySTDinterwordspacing

\bibitem{Latif2018AdversarialRobustness}
\BIBentryALTinterwordspacing
S.~Latif, R.~Rana, and J.~Qadir, ``{Adversarial Machine Learning And Speech
  Emotion Recognition: Utilizing Generative Adversarial Networks For
  Robustness},'' \emph{arXiv}, 11 2018. [Online]. Available:
  \url{http://arxiv.org/abs/1811.11402}
\BIBentrySTDinterwordspacing

\bibitem{Rana2019AutomatedFuture}
\BIBentryALTinterwordspacing
R.~Rana, S.~Latif, R.~Gururajan, A.~Gray, G.~Mackenzie, G.~Humphris, and
  J.~Dunn, ``{Automated screening for distress: A perspective for the
  future},'' \emph{European Journal of Cancer Care}, vol.~28, no.~4, 7 2019.
  [Online]. Available: \url{https://pubmed.ncbi.nlm.nih.gov/30883964/}
\BIBentrySTDinterwordspacing

\bibitem{Hester2017DeepDemonstrations}
\BIBentryALTinterwordspacing
T.~Hester, M.~Vecerik, O.~Pietquin, M.~Lanctot, T.~Schaul, B.~Piot, D.~Horgan,
  J.~Quan, A.~Sendonaris, G.~Dulac-Arnold, I.~Osband, J.~Agapiou, J.~Z. Leibo,
  and A.~Gruslys, ``{Deep Q-learning from Demonstrations},''
  \emph{arXiv:1704.03732 [cs]}, 2017. [Online]. Available:
  \url{http://arxiv.org/abs/1704.03732}
\BIBentrySTDinterwordspacing

\bibitem{Watkins1989LearningRewards}
C.~J. C.~H. Watkins, ``{Learning from Delayed Rewards},'' Ph.D. dissertation,
  Cambridge, UK, 1989.

\bibitem{Wiering1999ExplorationsLearning}
\BIBentryALTinterwordspacing
M.~Wiering, ``{Explorations in Efficient Reinforcement Learning},'' Ph.D.
  dissertation, 1999. [Online]. Available:
  \url{https://dare.uva.nl/search?identifier=6ac07651-85ee-4c7b-9cab-86eea5b818f4}
\BIBentrySTDinterwordspacing

\bibitem{Cesa-Bianchi2017BoltzmannRight}
N.~Cesa-Bianchi, C.~Gentile, G.~Lugosi, and G.~Neu, ``{Boltzmann exploration
  done right},'' ser. NIPS'17.\hskip 1em plus 0.5em minus 0.4em\relax Curran
  Associates Inc., 2017, pp. 6287--6296.

\bibitem{Pan2020ReinforcementUpdates}
\BIBentryALTinterwordspacing
L.~Pan, Q.~Cai, Q.~Meng, W.~Chen, and L.~Huang, ``{Reinforcement learning with
  dynamic boltzmann softmax updates},'' in \emph{IJCAI International Joint
  Conference on Artificial Intelligence}, vol. 2021-January.\hskip 1em plus
  0.5em minus 0.4em\relax International Joint Conferences on Artificial
  Intelligence, 7 2020, pp. 1992--1998. [Online]. Available:
  \url{https://www.ijcai.org/proceedings/2020/276}
\BIBentrySTDinterwordspacing

\bibitem{Leibfried2018Model-BasedNetworks}
\BIBentryALTinterwordspacing
F.~Leibfried and P.~Vrancx, ``{Model-Based Regularization for Deep
  Reinforcement Learning with Transcoder Networks},'' \emph{arXiv}, 9 2018.
  [Online]. Available: \url{http://arxiv.org/abs/1809.01906}
\BIBentrySTDinterwordspacing

\bibitem{ElAyadi2011SurveyDatabases}
M.~El~Ayadi, M.~S. Kamel, and F.~Karray, ``\BIBforeignlanguage{en}{{Survey on
  speech emotion recognition: Features, classification schemes, and
  databases}},'' \emph{\BIBforeignlanguage{en}{Pattern Recognition}}, vol.~44,
  no.~3, pp. 572--587, 2011.

\bibitem{Ververidis2006EmotionalMethods}
D.~Ververidis and C.~Kotropoulos, ``\BIBforeignlanguage{en}{{Emotional speech
  recognition: Resources, features, and methods}},''
  \emph{\BIBforeignlanguage{en}{Speech Communication}}, vol.~48, no.~9, pp.
  1162--1181, 2006.

\bibitem{Schuller2003HiddenRecognition}
B.~Schuller, G.~Rigoll, and M.~Lang, ``{Hidden Markov model-based speech
  emotion recognition},'' in \emph{2003 IEEE International Conference on
  Acoustics, Speech, and Signal Processing, 2003. Proceedings. (ICASSP '03).},
  vol.~2, 2003, pp. II--1.

\bibitem{Cairns1994NonlinearConditions}
\BIBentryALTinterwordspacing
D.~A. Cairns and J.~H.~L. Hansen, ``{Nonlinear analysis and classification of
  speech under stressed conditions},'' \emph{The Journal of the Acoustical
  Society of America}, vol.~96, no.~6, pp. 3392--3400, 1994. [Online].
  Available: \url{https://asa.scitation.org/doi/10.1121/1.410601
  files/674/1.html}
\BIBentrySTDinterwordspacing

\bibitem{Lee2002CombiningRecognition}
\BIBentryALTinterwordspacing
C.~Lee, S.~S. Narayanan, and R.~Pieraccini, ``{Combining acoustic and language
  information for emotion recognition},'' 2002. [Online]. Available:
  \url{https://www.semanticscholar.org/paper/Combining-acoustic-and-language-information-for-Lee-Narayanan/d5abf8adb874577dffc4038207b6b91bee0a3450}
\BIBentrySTDinterwordspacing

\bibitem{Latif2019DirectSpeech}
\BIBentryALTinterwordspacing
S.~Latif, R.~Rana, S.~Khalifa, R.~Jurdak, and J.~Epps, ``{Direct Modelling of
  Speech Emotion from Raw Speech},'' 2019. [Online]. Available:
  \url{http://arxiv.org/abs/1904.03833}
\BIBentrySTDinterwordspacing

\bibitem{Han2014SpeechMachine}
K.~Han, D.~Yu, and I.~Tashev, ``{Speech Emotion Recognition Using Deep Neural
  Network and Extreme Learning Machine},'' in \emph{Interspeech 2014}, 9 2014.

\bibitem{Latif2018CrossTechnique}
S.~Latif, R.~Rana, S.~Younis, J.~Qadir, and J.~Epps, ``{Cross Corpus Speech
  Emotion Classification - An Effective Transfer Learning Technique},'' 2018.

\bibitem{Huang2019SpeechSounds}
K.~Y. Huang, C.~H. Wu, Q.~B. Hong, M.~H. Su, and Y.~H. Chen, ``{Speech Emotion
  Recognition Using Deep Neural Network Considering Verbal and Nonverbal Speech
  Sounds},'' in \emph{ICASSP, IEEE International Conference on Acoustics,
  Speech and Signal Processing - Proceedings}, vol. 2019-May.\hskip 1em plus
  0.5em minus 0.4em\relax Institute of Electrical and Electronics Engineers
  Inc., 5 2019, pp. 5866--5870.

\bibitem{983933}
R.~{Polikar}, L.~{Upda}, S.~S. {Upda}, and V.~{Honavar}, ``Learn++: an
  incremental learning algorithm for supervised neural networks,'' \emph{IEEE
  Transactions on Systems, Man, and Cybernetics, Part C (Applications and
  Reviews)}, vol.~31, no.~4, pp. 497--508, 2001.

\bibitem{zhao2010incremental}
Q.~L. Zhao, Y.~H. Jiang, and M.~Xu, ``Incremental learning by heterogeneous
  bagging ensemble,'' in \emph{International Conference on Advanced Data Mining
  and Applications}.\hskip 1em plus 0.5em minus 0.4em\relax Springer, 2010, pp.
  1--12.

\bibitem{Busso2008IEMOCAP:Database}
\BIBentryALTinterwordspacing
C.~Busso, M.~Bulut, C.-C. Lee, A.~Kazemzadeh, E.~Mower, S.~Kim, J.~N. Chang,
  S.~Lee, and S.~S. Narayanan, ``\BIBforeignlanguage{en}{{IEMOCAP: interactive
  emotional dyadic motion capture database}},''
  \emph{\BIBforeignlanguage{en}{Language Resources and Evaluation}}, vol.~42,
  no.~4, p. 335, 2008. [Online]. Available:
  \url{https://doi.org/10.1007/s10579-008-9076-6}
\BIBentrySTDinterwordspacing

\bibitem{Haq2008Audio-visualClassification}
S.~Haq, P.~J.~B. Jackson, and J.~Edge, ``{Audio-visual feature selection and
  reduction for emotion classification},'' 2008.

\bibitem{Davis1980ComparisonSentences}
S.~Davis and P.~Mermelstein, ``{Comparison of parametric representations for
  monosyllabic word recognition in continuously spoken sentences},'' \emph{IEEE
  transactions on acoustics, speech, and signal processing}, vol.~28, no.~4,
  pp. 357--366, 1980.

\bibitem{McFee2015Librosa:Python}
B.~McFee, C.~Raffel, D.~Liang, D.~P.~W. Ellis, M.~McVicar, E.~Battenberg, and
  O.~Nieto, ``{librosa: Audio and music signal analysis in python},'' vol.~8,
  2015.

\bibitem{Chollet2015Keras}
\BIBentryALTinterwordspacing
F.~Chollet and {others}, \emph{{Keras}}, 2015. [Online]. Available:
  \url{https://keras.io}
\BIBentrySTDinterwordspacing

\bibitem{Abadi2015TensorFlow:Systems}
\BIBentryALTinterwordspacing
M.~Abadi, A.~Agarwal, P.~Barham, E.~Brevdo, Z.~Chen, C.~Citro, G.~S. Corrado,
  A.~Davis, J.~Dean, M.~Devin, S.~Ghemawat, I.~Goodfellow, A.~Harp, G.~Irving,
  M.~Isard, Y.~Jia, R.~Jozefowicz, L.~Kaiser, M.~Kudlur, J.~Levenberg,
  D.~Man{\'{e}}, R.~Monga, S.~Moore, D.~Murray, C.~Olah, M.~Schuster,
  J.~Shlens, B.~Steiner, I.~Sutskever, K.~Talwar, P.~Tucker, V.~Vanhoucke,
  V.~Vasudevan, F.~Vi{\'{e}}gas, O.~Vinyals, P.~Warden, M.~Wattenberg,
  M.~Wicke, Y.~Yu, and X.~Zheng, \emph{{TensorFlow: Large-Scale Machine
  Learning on Heterogeneous Systems}}, 2015. [Online]. Available:
  \url{https://www.tensorflow.org/}
\BIBentrySTDinterwordspacing

\bibitem{Latif2020DeepTrends}
\BIBentryALTinterwordspacing
S.~Latif, R.~Rana, S.~Khalifa, R.~Jurdak, J.~Qadir, and B.~W. Schuller, ``{Deep
  Representation Learning in Speech Processing: Challenges, Recent Advances,
  and Future Trends},'' \emph{arXiv:2001.00378 [cs, eess]}, 2020. [Online].
  Available: \url{http://arxiv.org/abs/2001.00378}
\BIBentrySTDinterwordspacing

\end{thebibliography}

\begin{IEEEbiography}
	[{\includegraphics[width=1in,height=1.25in,clip,keepaspectratio]{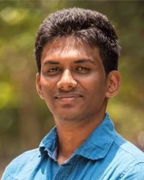}}]
	{Thejan Rajapakshe}
	 received bachelor degree in Applied Sciences from Rajarata University of Sri Lanka, Mihintale and Bachelor of Information Technology from University of Colombo, School of Computing in 2016. He is currently a research scholar at the University of Southern Queensland (USQ). He is also an Associate Technical Team Lead at CodeGen International - Research \& Development. His research interests include reinforcement learning, speech processing and deep learning.
\end{IEEEbiography}

\begin{IEEEbiography}
	[{\includegraphics[width=1in,height=1.25in,clip,keepaspectratio]{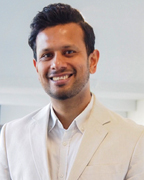}}]
	{Rajib Rana}
	is an experimental computer scientist, Advance Queensland Research Fellow and a Senior Lecturer in the University of Southern Queensland. He is also the Director of IoT Health research program at the University of Southern Queensland. He is the recipient of the prestigious Young Tall Poppy QLD Award 2018 as one of Queensland’s most outstanding scientists for achievements in the area of scientific research and communication. Rana’s research work aims to capitalise on advancements in technology along with sophisticated information and data processing to better understand disease progression in chronic health conditions and develop predictive algorithms for chronic diseases, such as mental illness and cancer. His current research focus is on Unsupervised Representation Learning. He received his B.Sc. degree in Computer Science and Engineering from Khulna University, Bangladesh with Prime Minister and President’s Gold medal for outstanding achievements and Ph.D. in Computer Science and Engineering from the University of New South Wales, Sydney, Australia in 2011. He received his postdoctoral training at Autonomous Systems Laboratory, CSIRO before joining the University of Southern Queensland as Faculty in 2015.
\end{IEEEbiography}

\begin{IEEEbiography}
	[{\includegraphics[width=1in,height=1.25in,clip,keepaspectratio]{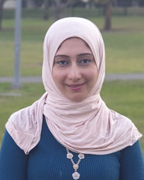}}]
	{Sara Khalifa}
	is currently a senior research scientist at the Distributed Sensing Systems research group, Data61—CSIRO. She is also an honorary adjunct lecturer at University of Queensland and conjoint lecturer at University of New South Wales.  Her research interests rotate around the broad aspects of mobile and ubiquitous computing, mobile sensing and Internet of Things (IoT). She obtained a PhD in Computer Science and Engineering from UNSW (Sydney, Australia). Her PhD dissertation received the 2017 John Makepeace Bennett Award which is awarded by CORE (Computing Research and Education Association of Australasia) to the best PhD dissertation of the year within Australia and New Zealand in the field of Computer Science. Her research has been recognised by multiple iAwards including 2017 NSW Mobility Innovation of the year, 2017 NSW R\&D Innovation of the year, National Merit R\&D Innovation of the year, and the Merit R\&D award at the Asia Pacific ICT Alliance (APICTA) Awards, commonly known as the "Oscar" of the ICT industry in the Asia Pacific, among others.
\end{IEEEbiography}

\begin{IEEEbiography}
	[{\includegraphics[width=1in,height=1.25in,clip,keepaspectratio]{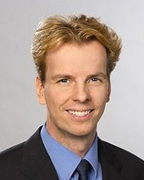}}]
	{Bj\"{o}rn W.\ Schuller}
	received his diploma in 1999, his doctoral degree for his study on Automatic Speech and Emotion Recognition in 2006, and his habilitation and Adjunct Teaching Professorship in the subject area of Signal Processing and Machine Intelligence in 2012, all in electrical engineering and information technology from TUM in Munich/Germany. He is Professor of Artificial Intelligence in the Department of Computing at the Imperial College London/UK, where he heads GLAM — the Group on Language, Audio \& Music, Full Professor and head of the ZD.B Chair of Embedded Intelligence for Health Care and Wellbeing at the University of Augsburg/Germany, and CEO of audEERING. He was previously full professor and head of the Chair of Complex and Intelligent Systems at the University of Passau/Germany. Professor Schuller is Fellow of the IEEE, Golden Core Member of the IEEE Computer Society, Senior Member of the ACM, President-emeritus of the Association for the Advancement of Affective Computing (AAAC), and was elected member of the IEEE Speech and Language Processing Technical Committee. He (co-)authored 5 books and more than 800 publications in peer-reviewed books, journals, and conference proceedings leading to more than overall 25 000 citations (h-index = 73). Schuller is general chair of ACII 2019, co-Program Chair of Interspeech 2019 and ICMI 2019, repeated Area Chair of ICASSP, and former Editor in Chief of the IEEE Transactions on Affective Computing next to a multitude of further Associate and Guest Editor roles and functions in Technical and Organisational Committees.
\end{IEEEbiography}

\begin{IEEEbiography}
	[{\includegraphics[width=1in,height=1.25in,clip,keepaspectratio]{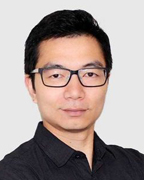}}]
	{Jiajun Liu}
	is a science leader at CSIRO, Australia. His current research interests include multimedia content analysis, indexing, and retrieval. He received the B.E. degree from Nanjing University, China, and the Ph.D. degree from the University of Queensland, Brisbane, QLD, Australia, in 2013. He was also a Researcher/Software Engineer for IBM, China, during 2006 to 2008.
\end{IEEEbiography}

\end{document}